\documentstyle[emulateapj,apjfonts,onecolfloat]{article}

\begin{document}

\twocolumn[
\submitted{ApJ submitted}

\lefthead{CONSTRAINING \mbox{$q_0$} WITH CLUSTER GAS MASS FRACTIONS: A
FEASIBILITY STUDY}
\righthead{RINES ET AL.}

\title{CONSTRAINING \lowercase{$q_0$} WITH CLUSTER GAS MASS FRACTIONS: A
FEASIBILITY STUDY}

\author{K. Rines, W. Forman, U. Pen, C. Jones}
\date{July 1998}
\affil{Harvard-Smithsonian Center for Astrophysics, 60 Garden St., 
Cambridge, MA 02138; krines,wforman,upen,cjones@cfa.harvard.edu}
\author{R. Burg}
\affil{Johns Hopkins Univ. Physics \& Astronomy Dept.,
The Bloomberg Center, 
Baltimore, MD 21218 ; burg@stsci.edu}
\authoraddr{krines@cfa.harvard.edu}

\begin{abstract}
As the largest gravitationally bound objects in
the universe, clusters of galaxies may contain a fair 
sample of the baryonic mass fraction of the universe. 
Since the gas mass fraction from the hot ICM is believed to be constant
in time, the value of the cosmological deceleration parameter $q_0$
can be determined, in principle, by comparing the calculated gas mass
fraction in nearby and distant clusters (Pen 1997). To test the
potential of this method, we compare the gas fractions
derived for a sample of luminous ($L_X > 10^{45}$erg 
s$^{-1}$), nearby clusters with those calculated for eight luminous,
distant ($0.3 < z < 0.6$) clusters using 
$ASCA$ and $ROSAT$ observations. For consistency, we
evaluate the gas mass fraction at a fixed physical radius of 1
$h_{50}^{-1}$ Mpc (assuming $q_0$=0.0).
We find a best fit value of $q_0 = 0.07$ with -0.47 $< q_0 < $ 0.67 at 95\%
confidence. This analysis includes both measurement errors and an
intrinsic 25\% scatter in the gas fractions due to the effects of
cooling flows and mergers.
We also determine the gas fraction using the method of
Evrard, Metzler, \& Navarro (1997) 
to find the total mass within $r_{500}$, the radius where the 
mean overdensity of matter is 500 times the critical density. 
In simulations, this method reduces the scatter in the determination
of gravitational mass without biasing the mean.
We find that it also reduces the scatter in actual
observations for nearby clusters, but not as much as
simulations suggest. Using this method, the best fit value is $q_0 =
0.04$ with 
-0.50 $ < q_0 <$ 0.64. The excellent agreement between these two
methods suggests that this may be a useful technique for determining
$q_0$. The constraints on $q_0$ should improve as more distant
clusters are studied and precise temperature profiles are measured to
large radii.

\end{abstract}
]

\section{INTRODUCTION}	
One of the major goals of cosmology has been to measure accurately the
cosmological deceleration parameter $q_0$. Determining
$q_0$ from the count magnitude relation for galaxies is
notoriously difficult due to evolutionary effects (e.g., Sandage
1988). However, recent studies of galaxy counts from redshift surveys
provide reasonable constraints on $q_0$. Schuecker et al. (1998) find
a best fit of $q_0$ = 0.1 with an upper
limit of $q_0 <$ 0.75 at 95\% confidence assuming $\Lambda = 0$. 
Studies of high-redshift Type Ia supernovae (Perlmutter et al. 1997,
Garnavich et al. 1998, Riess et al. 1998) suggest that $q_0$ may be
negative due to a non-zero cosmological constant. Finally, a study
of gravitational lensing statistics suggests that $q_0 > -2$ (Park \&
Gott 1998). These constraints do not, however, allow us to rule out
any of the most popular cosmological models (e.g., closed universes
with $q_0 > 0.5$ or $\Omega + \Lambda = 1$ universes with $q_0 \sim
-0.5$). It has been proposed that a constant $L_X-T$ relation would
allow cosmological parameters to be measured (Henry \& Tucker 1979).
Another interesting method for improving or reinforcing these
constraints is to analyze the gas mass fractions of rich galaxy
clusters. Since 
this gas mass fraction is likely to be constant and its calculated
value depends on cosmological parameters, comparisons of gas mass
fractions of nearby and distant clusters can constrain $q_0$ (Sasaki
1996, Pen 1997). In this paper, we test the possibility of using the
constancy of the gas mass fraction to constrain $q_0$. We use the
available data with suitable analyses to evaluate plausible
uncertainties to determine if the method is sufficiently sensitive to
allow accurate measurements of $q_0$ using future {\mbox AXAF} and
{\mbox XMM} observations of large cluster samples.

The approach we follow to constrain $q_0$ requires that the mean
cluster gas mass fraction be a universal constant, independent of
redshift. Rich clusters have masses $\sim 10^{15}
M_\odot$, which suggests that the matter they contain
ought to be a fair sample of the distribution of baryons and
dark matter in the universe. Detailed analyses for hot, massive
clusters (White et al. 1993, Evrard, Metzler, \& Navarro 1996;
hereafter EMN) 
show that there is no known mechanism to significantly concentrate baryons
(both galaxies and X-ray gas)
in the center of the cluster or to expel them from the
potential. This is supported by numerical simulations
(e.g., White et al. 1993, Lubin et al. 1996), which show that the X-ray
gas mass fraction $f_g = M_{gas}/M_{tot}$  
averaged within $\sim 1$ Mpc should be constant, independent
of redshift or evolution (Danos \& Pen 1998, Frenk et al. 
1998). In a flat universe, there is a degeneracy between
redshift and $\sigma_8$, or equivalently, epoch and mass. Evrard (1997)
has shown that the gas mass fraction is independent of total cluster
mass to an accuracy of 1.5\%. This is equivalent to an independence of
the gas mass fraction in
redshift to this accuracy. Alternatively, in an extremely open
universe, clusters formation ceases early in the evolution of the
universe. Afterwards, there is little evolution imposed on clusters
by their environments. 
Most simulations (Thomas \& Couchmann 1992, Katz, Hernquist, \&
Weinberg 1992, Cen \& Ostriker 1993, Kang et al. 1994, Lubin et
al. 1996, Pen 1998) show the gas to be slightly more extended than
the dark matter within $\sim 1$ Mpc due to shock heating and other
transfers of energy. This raises a possible concern, namely that the gas
fraction may depend on radius. At the virial radius, inflow and
outflow of mass will affect hot gas and dark matter identically, so
that the gas mass fraction should remain constant at the virial
radius. However, since the X-ray data do not extend to the virial
radius, we determine the gas mass fraction at two radii -- a fixed
physical radius of 1 $h_{50}^{-1}$ Mpc and a fixed overdensity radius
$r_{500}$ (described below). If there are any large systematic
variations of gas mass fraction with radius, these two methods
should yield significantly different values of $q_0$. As described
below, they agree remarkably well. 

Evolutionary effects due to stripping of gas from galaxies and
ejection of mass by supernovae should be relatively small. 
In nearby late-type galaxies,
neutral gas contains $\sim 10\%$ of the luminous stellar mass. Since
mass-to-light ratios of clusters are typically 300 $h_{50}^{-1}$, the
gas associated with the galaxies in a rich cluster comprises $\sim 3
\times 10^{-4} h_{50}^{-1}$ of the total cluster mass. Since X-ray gas
mass fractions are typically $0.18 h_{50}^{-3/2}$, this implies that
the ratio of galactic gas to X-ray gas is $\sim 0.002
h_{50}^{1/2}$. For early-type galaxies, Faber \& Gallagher (1976) find
that the amount of gas ejected into the ISM is 0.015 $M_\odot
\mbox{yr}^{-1} (10^9 L_\odot)^{-1}$. For 1000 bright galaxies
($L\sim10^{11}L_\odot$), this amounts to $\sim 2 \times 10^{12}
M_\odot$ over a Hubble time ($10^{10}$ yrs) for the entire
cluster. Since the mass in hot gas within the central 1 $h_{50}^{-1}$
Mpc is $\sim10^{14} M_\odot$ for the clusters in the current study,
this effect amounts to less than 2\% of the hot gas mass in the core
of the cluster. Thus, the effect of gas stripping from galaxies,
both for late-type and early-type dominated populations, is almost
certainly less than one percent of the mass in X-ray gas.

Similarly, the ratio of luminous stellar mass to X-ray
gas mass in clusters is $0.06 h_{50}^{-1}$. Since only stars larger
than about $6 M_\odot$ become supernovae, and since these massive
stars comprise a small fraction of luminous stellar mass, any
injection of gas into the ICM from supernovae will be an effect of
less than one percent. Thus, the actual gas mass fraction should vary
by no more than two percent due to the effects of gas stripping of
galaxies and ejection of gas from galaxies into the ICM by supernovae. 
Since the true gas mass fraction is assumed to remain constant while the
calculated gas mass fraction depends on  
cosmological parameters, any apparent change in the cluster gas mass
fraction with redshift can be used
to directly measure the evolution of the angular diameter distance 
(Sasaki 1996, Pen 1997). As studied here, the cosmological 
dependence of $f_g$ is roughly $f_g \propto d_L^{3/2}$, where $d_L$ is
the luminosity distance. The cluster gas mass fraction method provides a 
direct measurement of $q_0$ (Pen 1997). The difference between a flat
universe and an empty universe at a redshift of $\sim 0.5$ is 15\%,
significantly larger than the systematic effects described above.

There have been some attempts to determine cosmological parameters 
using gas fractions from the literature (Sasaki 1996,
Pen 1997, Danos \& Pen 1998, Cooray 1998), but the gas fractions for
distant clusters are found 
by different authors with unknown systematic differences in the
measured fractions depending on the methods of data analysis. 
These studies also rely on small numbers of high redshift clusters. Since
simulations predict an intrinsic scatter in the observed gas fraction
of 25\% (Pen 1998, Danos \& Pen 1998), a large sample is needed to
avoid excessive noise due to small sample size. We present here
the first ``large'' sample of cluster gas fractions at high
redshifts determined
using an internally consistent method of analysis. This study is also
the first to select nearby clusters on the basis of
X-ray luminosity, which is known to be strongly correlated with other
cluster properties, most notably the gas temperature
(e.g., David et al. 1993). Our sample consists of eight clusters at $z
\gtrsim 0.3$, which we compare to a sample  
of nearby clusters with comparable luminosities ($L_X > 10^{45}$erg
s$^{-1}$). These are all of the distant ($z \gtrsim 0.3$) clusters
available in both the $ASCA$ public archive and the $ROSAT$ public
archive which were observed with the $PSPC$ and can be well-described
by the standard hydrostatic-isothermal $\beta$ model.
Since the cosmological effects depend only on the apparent change of
gas fraction, they are independent of the numerical 
value of the gas fraction. Thus, systematic uncertainties in the
numerical value of the gas fraction (e.g., from uncertainties in $H_0$)
are unimportant provided they are consistently applied to all 
clusters in the sample. 

To determine the total gas mass and total gravitational mass
in a cluster, we must determine
its temperature, temperature distribution, and gas density distribution. 
$ASCA$ has a large energy bandwidth (0.5-10.0 keV), making it
especially useful for measuring cluster temperatures, which are
typically in the range 1-10 keV.  
However, $ASCA$ has insufficient spatial resolution to accurately
determine the gas distribution, so we must use $ROSAT$,
with its higher angular resolution but lower energy bandwidth, to derive
the gas distributions. As described below, we obtain -0.47 $< q_0 < $
0.74 and -0.50 $ < q_0 <$ 0.64 at 95\% confidence using two
similar techniques for calculating the gas fractions. We describe
the techniques for calculating gas fractions in Section 2 and the
observations in Section 3. We
calculate the cosmological dependence of the gas fraction and $q_0$ in
Section 4 and discuss our results in Section 5.

\section{Methods of Calculating Gas Mass Fractions}

Since the dynamical time of the gas is
much shorter than the age of the cluster, it is 
reasonable to assume that the gas is in hydrostatic
equilibrium near the center of the cluster.
Indeed, Jones \& Forman (1984, 1998, hereafter JF98) found that a
spherically symmetric hydrostatic-isothermal 
$\beta$ model (Cavaliere \& Fusco-Femiano 1976)
is a surprisingly accurate description of cluster profiles. This model
provides good fits to the present data, so we use it in our analysis. 
The surface brightness profile, $\Sigma (r)$, of a cluster can be  
accurately described as a function of radius $r$ by a $\beta$ model 
\begin{equation}
\Sigma(r) = \Sigma_0 \Bigl(1 + (\frac{r}{a})^2 \Bigr)^{-3\beta + 1/2}
\end{equation}
where $a$ is the core radius of the cluster, $\beta = \mu m_p
\sigma ^2/3kT_{gas}$ is the ratio 
of energy per unit mass in galaxies to the energy per unit mass in
gas where $\sigma$ is the velocity dispersion, $\mu$ is the mean
molecular weight, and $T_{gas}$ (hereafter
T or $T_X$) is the
temperature of the hot gas (Forman \& Jones
1982). $\beta$ has been  found to have a numerical value of $0.6 \pm
0.1$ for 60 of 85 clusters (Jones \& Forman 1998). This is different
than the value of $\beta \sim 0.8$ found in some simulations (Evrard
1990), but this apparent discrepancy was resolved by Navarro, Frenk,
\& White (1995) as an effect of observational constraints. They
further argue that the cause of the other ``beta-discrepancy'', the
difference between the value of $\beta_{spec} \sim 0.9$ measured by
spectroscopy of galaxies (e.g., Bahcall \& Lubin 1994) and the above
value is due to galaxies not being distributed like
an isothermal sphere, as is assumed by the model. Recent $ROSAT$
observations of nearby clusters suggest that the surface brightness
distribution of clusters is indeed accurately described by a $\beta$
model even out to large radii (Buote \& Canizares 1996).
For an isothermal gas, this yields a gas density profile described by
\begin{equation}
\label{density}
\rho_{gas}(r) = \rho_0(1 + (\case{r}{a})^2)^{-3\beta /2}
\end{equation}
where $r$ here (and for the rest of the paper) refers to the
three-dimensional radius. The central density can be found from the
luminosity $L(R)$ within a projected radius $R$ by the relation  
\begin{equation}
\label{cendensity}
L(R) = \frac {2 \pi n_e n_H \Lambda_0 a^3}{(1 - 3 \beta)}\int
_0^{\infty} \Bigl[ \Bigl( 1 + s^2 + \Bigl( \frac{R}{a}\Bigr) ^2
\Bigr) ^{-3\beta +1}- (1 + s^2)^{-3 \beta + 1}  \Bigr] ds
\end{equation}
where $\Lambda_0$ is the radiative cooling coefficient at the
gas temperature, $L(R)$ is the luminosity inside a radius $R$, $s$ is the
distance along the line of sight in units of core radius, and $n_e$
and $n_H$ are the electron and hydrogen number densities of the ICM
at the center of the cluster (e.g., David et al. 1990). The total gas
mass within a radius $r$ is 
then just the integral of the density distribution
(Equation~\ref{density}). For $ROSAT$ observations of hot clusters,
David et al. (1990) find that the gas mass is fairly insensitive to a
temperature gradient $T(r)$. 

With the assumptions of hydrostatic equilibrium and spherical symmetry,
the gravitational mass inside a radius r is given by
\begin{equation}
\label{mass1}
M_{tot}(<r) = - \frac{kT}{\mu m_p G} \Bigl({{d\ln \rho_{gas}} \over
{d\ln r}} + {{d\ln T} \over {d\ln r}} \Bigr) r
\end{equation}
Assuming a uniform temperature distribution, the second
term on the right hand side vanishes. We then only need to
determine the gas temperature and the density distribution of the gas
to calculate the gravitational mass. 
Combining equations~\ref{density} and~\ref{mass1}, 
we find that the mass is related to $\beta$ and
$a$ by 
\begin{equation}
\label{mass2}
M_{tot}(<r) = \frac{3kT\beta r^3}{\mu m_p Ga^2(1 + (\case{r}{a})^2)}
\newline = 1.13\times 10^{14} \beta T_{\mbox{keV}} \frac{r^3}{a^2 +
 r^2} M_{\sun}
\end{equation}
where $M_{tot}(<r)$ is the total gravitational mass within a radius
$r$ and the numerical approximation is valid for $T_{\mbox{keV}}$ in
keV and $r$ and $a$ in 
Mpc. The gas fraction $f_g(r)$ is then simply $M_{gas}/M_{tot}$
evaluated at a given radius.

The assumptions of hydrostatic equilibrium, spherical symmetry, and
isothermality are not perfect. We know that the latter two assumptions
are violated for some nearby clusters (Markevitch et al. 1998, Henry
\& Briel 1995), and clusters undergoing major mergers violate
hydrostatic equilibrium. In fact, Markevitch et al. suggest that all
clusters have temperature structure. The
effect of these violations of spherical symmetry, isothermality, and
hydrostatic equilibrium is to introduce an additional scatter in
the observed gas fractions beyond simple measurement errors. This
scatter has been found by simulations (Danos \& Pen 1998) to be about
25\%. We include this intrinsic scatter in our determinations of
$q_0$ in Section 4. 

Markevitch \& Vikhlinin (1997) studied the
variation of gas fraction with radius in detail for A2256. If A2256 is
assumed to be isothermal, the gas fraction varies by only about 10\%
with radius. However, by
including the temperature profile of the cluster, they found
that the isothermal model overestimated $f_g$ at small radii and
underestimated it at large radii, with the estimate being very close
to accurate at a radius of 1.2 $h_{50}^{-1}$ Mpc. 
We assume an isothermal model with a radius of 1 $h_{50}^{-1}$ Mpc, which
overestimates the gas fraction by about 15\% depending on the model
used for the dark matter distribution. 
Similar results are obtained for a study
of temperature profiles of 26 nearby clusters (Markevitch et
al. 1998). More importantly, the shapes of the temperature profiles
for these clusters was found to be extremely similar. This
similarity of cluster temperature profiles suggests that
cluster-to-cluster variations in temperature profile will have only a
small effect on the apparent change in gas fraction with redshift. 
Thus, since we lack detailed temperature profiles for distant
clusters, one of our approaches is to assume isothermality and
evaluate the gas fraction at 1 $h_{50}^{-1}$ Mpc.

The measurement of cluster gas fractions was explored in detail by
EMN (1996) based on cosmological gas dynamic simulations for several
cosmological models. They find that
cluster mass estimates based on the above method are surprisingly
unbiased, yielding an average estimated-to-true mass ratio of 1.02
with a standard deviation of 14\%-29\%. EMN also develop a method to
reduce the scatter in the estimate of cluster mass based solely on the
emission-weighted gas temperature $T_X$ within a radius where the mean
density is 500 times the critical density. This radius, denoted by
$r_{500}$, is found to vary with $T_X$ as 
\begin{equation}
r_{500} = (2.48 \pm 0.17) \Bigl( \frac{T_X}{10  \mbox{keV}} \Bigr)^{1/2}
h_{50}^{-1} \mbox{Mpc} .
\end{equation}
The mass within $r_{500}$ can be estimated as 
\begin{equation}
\label{massev}
M_{500}(T_X) = 2.22 \times 10^{15} \Bigl(\frac{T_X}{10  \mbox{keV}} \Bigr)
^{3/2} M_\odot
\end{equation}
and is found to have an average estimated-to-true mass ratio of 1.00
with a standard deviation of only 8\%-15\%. 
In Section 4, we use this method combined with the above method of
calculating the gas mass in an attempt to reduce the scatter in the
gas fraction and hence improve the constraints on $q_0$. For A2256
(Markevitch \& Vikhlinin 1997), the gas fraction at 
$r_{500}$ is underestimated by 20-40\% depending on the dark matter
distribution model. Again, this should mainly affect the absolute 
value of the gas fraction, not the apparent change in gas fraction as
a function of redshift and thus should not significantly affect our
results beyond contributing to the intrinsic scatter discussed above. 

Our use of two methods to determine the gas fraction can identify
possible systematic effects due to changes in gas fraction with
radius from observations. By comparing determinations of $q_0$ from
cluster gas mass 
fractions evaluated both at a fixed physical radius of 1 $h_{50}^{-1}$
Mpc and at a fixed overdensity radius ($r_{500}$), we should be able
to identify large systematic effects caused by our inability to
evaluate the gas mass fraction at the virial radius. 

\section{Observations}

$ASCA$'s broad energy band (0.5-10.0 keV) is particularly useful for 
determining hot ($>$ 5 keV) cluster temperatures. 
We therefore used $ASCA$ data to measure cluster temperatures and
luminosities, while
we used the higher angular resolution of $ROSAT$ to determine the
parameters of the gas distribution. Because of the poor angular
resolution of $ASCA$, we determine the emission weighted average
temperature within $6'$, which corresponds to  $\sim 2 h_{50}^{-1}$
Mpc at $z$=0.3. 

The temperatures and iron abundances of our sample are
shown in Table 1. Details of the fitting techniques can be found in
Rines et al. (1998).
The temperatures we measure are consistent with those found by
Mushotzky \& Scharf (1997) for clusters common to both samples.
The flux is measured between 0.01 and 100.0 keV from
the best-fit model, so it is effectively the bolometric flux.
The bolometric luminosity is calculated from the bolometric flux of
the best-fit model to the $ASCA$ data,
using the luminosity distance $d_L = c H_0^{-1}[z + \case{1}{2}(1 -
q_0)z^2]$ and assuming $q_0$ = 0.0.  

\begin{table*}[th] 
\begin{center}

\caption{\label{table1}}
{\sc Temperatures and iron abundances of several high redshift
clusters from $ASCA$ data.\\} 
\begin{tabular}{|lcccccccrc|}
\hline
\hline

Cluster& $z$ &$k$T & 90\% & Abund. & 90\% & Flux ($10^{-12}$) &
$L_X^{45}$(bol)\tablenotemark{a} &
\multicolumn{1}{c}{$\chi_\nu^2$} & Counts\tablenotemark{b}\\
 & & (keV) & (keV) & & & ergs cm$^{-2}$s$^{-1}$&
 ergs s$^{-1}$ & $\nu \simeq 1550$ &
(thousands) \\
\hline 
Zw3146 & 0.291 & 7.1 & 6.7-7.5 & 0.23 &	0.18-0.28 & 16.8 & 7.99 & 1.006 & 37.6\\
A1576 & 0.302 & 8.7 & 7.5-10.1 & 0.16 &	0.04-0.28 & 6.2 & 3.22 & 0.946 & 11.7\\
A2744 & 0.308 & 9.5 & 8.8-10.6 & 0.20 & 0.11-0.29 & 12.3 & 6.79 &
0.983 & 21.8\\
MS2137.3-2353 & 0.313 & 5.3 & 4.8-5.8 & 0.40 & 0.29-0.51 & 4.9 & 2.76
& 0.948 & 7.2\\
MS1358.4+6245 & 0.328 & 5.3 & 4.7-6.0	& 0.20	&0.09-0.32& 3.2	&
2.00 & 1.001 & 7.5 \\
A959 & 0.353 & 5.3 & 4.4-6.4 & 0.05  & $<$0.28 & 2.7 & 2.00 &0.977 & 4.6\\
MS0451.6-0305 & 0.541 & 9.0 & 7.5-10.8 & 0.19 & 0.07-0.33 & 3.2 & 6.46 & 0.975& 5.9\\
CL0016+16 & 0.555 & 8.6 & 7.3-10.2 & 0.11 & $<$0.22 & 4.0 & 8.56 &
0.976& 6.9\\
\hline
\end{tabular}
\end{center}
\tablenotetext{a}{Luminosities are calculated assuming $q_0$ = 0.0.}
\tablenotetext{b}{Total number of counts for all four detectors.}

\end{table*}

We used $ROSAT$ PSPC observations for analyzing surface brightness
profiles. Although the PSPC has less angular resolution than the HRI,
its background is much lower, which is important for extended faint
objects such as clusters. To account for the point spread
function of the PSPC, which depends on the angle from the center of the
instrument, we calculated the point spread function for each cluster
depending on its location in the field of view (Markevitch \&
Vikhlinin 1997). The surface brightness distribution was
fit to a $\beta$ model convolved with the PSF with $\beta$,
$\Sigma_0$, $a$, and a constant background as free parameters using
the method of 
maximum-likelihood. We exclude from our sample all clusters which
cannot be fit to a physically reasonable $\beta$ model (e.g., A851 and
A1758, which are known to have significant substructure -- see
Schindler \& Wambsganss 1996, Rines et al. 1998). 
The fit for A1576 is shown in Figure 1, and the resulting parameters
for all clusters are given in Table 2, along with the resulting gas
masses, gravitational masses, and gas fractions as calculated
assuming $q_0$ = 0.0. The parameters for CL0016+16 agree with those
found by Neumann \& B\"{o}hringer (1997) for HRI observations. 
Zw3146 is known to contain a large cooling
flow (Allen et al. 1996), so the temperature and hence the total mass
is likely underestimated. The observed gas fraction for this cluster 
would then be overestimated. With improved angular resolution, we
could exclude the cooling flow region and determine a more accurate
temperature for the cluster.

\begin{figure} [tb]
\plotone{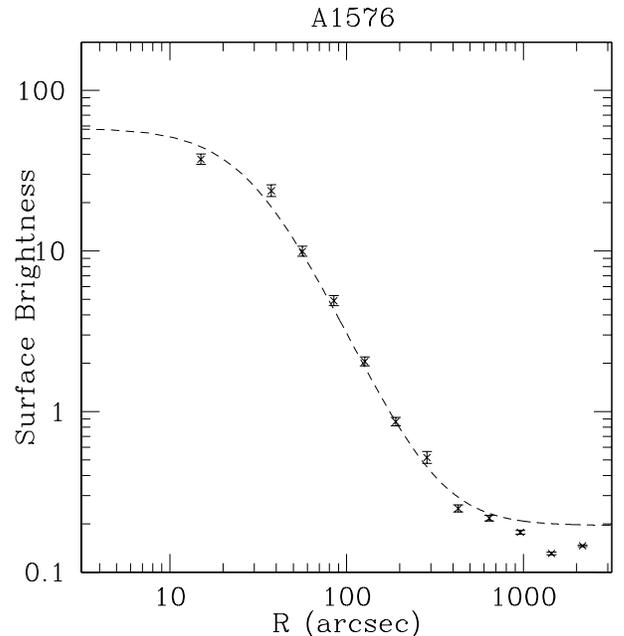}
\vskip -10pt

\figcaption{Surface brightness profile for A1576 and the best-fit $\beta$
model plus background. Errors shown are rms pixel-to-pixel variations.
}
\end{figure}

\begin{table*}[th]
\begin{center}

\caption{\label{table2}}
{\sc Gas distribution parameters from $ROSAT$ surface brightness
profiles.\\}
\begin{tabular}{|l|c|cc|c|ccc|ccc|}
\hline
\hline

Cluster& $z$ & $\beta$ & $a$ & $r_{500}$ & Gas Mass & Total Mass &
$f_g$ & Gas Mass & Total Mass & $f_g$ \\
 & & & ($h_{50}^{-1}$ Mpc)&($h_{50}^{-1}$ Mpc) & (10$^{14} M_\odot$)&(10$^{14} M_\odot$) & &(10$^{14} M_\odot$)
 & (10$^{14} M_\odot$) & \\
 & & & & &(1 $h_{50}^{-1}$ Mpc) &(1$h_{50}^{-1}$ Mpc) &(1$h_{50}^{-1}$ Mpc) &($r_{500}$) &($r_{500}$) &($r_{500}$) \\
\hline 
Zw3146 & 0.291 & 0.617 & 0.079 &2.09 & 1.25 & 5.05 & 0.247 & 2.97  &
13.3 & 0.224 \\
A1576 & 0.302 & 0.566 & 0.179 & 2.31 & 0.908 & 5.39 & 0.169 & 3.04 & 18.0 &
0.169 \\
A2744 & 0.308 & 0.887 & 0.654 & 2.42 & 1.49 & 6.67 & 0.224 & 4.85 &
20.6 & 0.236 \\
MS2137 & 0.313 & 0.642 & 0.054 & 1.81 & 0.642& 3.83 & 0.167 & 1.25 &
8.57 & 0.146 \\
MS1358 & 0.328 & 0.641 & 0.112 & 1.81 &0.700& 3.79 & 0.185 & 1.42 &
8.57 & 0.166 \\
A959  & 0.353 & 0.732 & 0.563 &1.81 &0.858 & 3.33 & 0.258 & 2.17 & 8.57 &
0.253 \\
MS0451 & 0.541 & 0.929 & 0.335 & 2.35 & 1.32 & 8.47 & 0.156 & 2.78 &
18.9 & 0.147 \\
CL0016 & 0.555 & 0.869 & 0.427 & 2.30 & 1.67 & 7.14 & 0.234 & 4.19 & 17.7 &
0.236 \\

\hline
\end{tabular}
\end{center}
NOTE.---%
Masses calculated assuming $q_0$=0.0.
\end{table*}

Table 2 shows the gas fractions both at 1 $h_{50}^{-1}$ Mpc and at
$r_{500}$, and we compare these gas fractions in Figure 3. 
These two methods use equations~\ref{mass2} and~\ref{massev}
respectively to estimate $M_{tot}$; both use equations~\ref{density}
and~\ref{cendensity} to estimate the gas mass. The values
of $f_g(r_{500})$ are lower on average than $f_g(1 h_{50}^{-1}$ Mpc),
but the difference lies mostly within the fractional errors taken
from the 90\% confidence limits on the temperature. Evrard (1997)
finds that in simulations, the gas fraction at $r_{500}$ is related to
the gas fraction at $r_X$,
according to $f_g(r_{500}) = f_g(r_X)\times (r_{500}/r_X)^{\eta}$,
where $\eta$ = 0.13-0.17. The above discussion of the effect of
temperature profiles on $f_g$ suggests that $f_g(r_{500})$ is
understimated by assuming isothermality. This effect is in the
sense of reducing the difference between the simulations of Evrard
and the present results.

\placetable{table1}

\section{Comparison to Nearby Clusters and Constraints on \lowercase{$q_0$}}

To determine the apparent change of the measured gas fraction, we
must know the gas fraction at the present epoch. We do this by
averaging the gas fractions found by JF98 for clusters observed with
{\it Einstein} at $z < 0.1$ with luminosities greater 
than 10$^{45}$ erg s$^{-1}$ and accurate measurements
of $\beta$ and core radius $a$. Comparing {\it Einstein} to $ASCA$
data introduces only a small error; Henry (1997) found that the fluxes
determined from $ASCA$ for $EMSS$ clusters agree very well with those
found by {\it Einstein}. For ten distant clusters, the mean flux ratio
is 0.998, with an rms scatter of 0.205.
There is excellent agreement between the gas fractions
calculated from the method of equations~\ref{density}
and~\ref{cendensity} and those in JF98, so the relative gas fractions
of nearby versus distant clusters should be accurate. 
The luminosity selection 
ensures that the nearby clusters are similar to those we study at high
redshift. For example, the well-known relation between X-ray
luminosity and temperature (e.g., David et al. 1993, Mushotzky \&
Scharf 1997) indicates that clusters with comparable luminosities  
also have comparable gas temperatures. 
It has been found that the physical properties of individual
clusters such as iron abundance, luminosity-temperature relation,
and X-ray temperature do not evolve significantly for redshifts up to
$z \sim 0.4$ (Mushotzky \& Loewenstein 1997, Mushotzky \& Scharf
1997). Thus, both simulations and observations suggest
that the dependence of the gas fraction on cluster
evolution should be small for our sample. Finally, it is worth noting
that while the above luminosity selection ensures that the two samples
of clusters are similar, we are not required by the method to make
such a selection. For instance, Evrard (1997) showed that the gas mass
fraction should be a universal constant for rich clusters and even
groups of galaxies. However, selecting our nearby sample by luminosity
removes a possible systematic effect (though we shall see below that
it makes very little difference in the results).

\begin{table*}[th]

\caption{\label{table3}}
\begin{center}
{\sc Gas and gravitational masses and gas fractions for
luminous nearby clusters (JF98) using the EMN method.\\}
\begin{tabular}{|lcccccccc|}
\hline
\hline

Cluster & $z$ & $\beta$ & $a$ & $T_X$ & $r_{500}$ & Gas Mass & Total
Mass & $f_g$  \\
  & & & ($h_{50}^{-1}$ Mpc) & (keV) & ($h_{50}^{-1}$ Mpc)& 
(10$^{14} M_\odot$)& (10$^{14} M_\odot$) &   \\
\hline 
 A85    &0.0518	&0.62 	& 0.26  & 6.2	& 1.95	& 2.03	& 10.7	& 0.189	\\
 A401	&0.0748	&0.65	& 0.26	& 7.8	& 2.19	& 2.46 	& 15.3	& 0.161	\\
 A426	&0.0183	&0.58	& 0.28	& 5.9	& 1.90	& 2.85	& 10.1	& 0.283	\\
 A478	&0.0881	&0.76	& 0.30	& 7.3	& 2.12	& 2.79	& 13.9	& 0.202	\\
 A3266	&0.0594	&0.70	& 0.55	& 6.2	& 1.95	& 2.39	& 10.8	& 0.221	\\
 A644	&0.0704	&0.70	& 0.20	& 6.9	& 2.06	& 1.53	& 12.7	& 0.120	\\
 A754	&0.0528	&0.80	& 0.58	& 9.1	& 2.37	& 2.65	& 19.3	& 0.137	\\
 A1650	&0.0845	&0.78	& 0.29	& 5.5	& 1.84	& 1.35	& 9.06	& 0.149	\\
 A1795	&0.0616	&0.73	& 0.29	& 5.6	& 1.86	& 1.72	& 9.30	& 0.185	\\
 A2029	&0.0767	&0.69	& 0.20	& 7.8	& 2.19	& 2.50	& 15.3	& 0.164	\\
 A2065	&0.0721	&0.64	& 0.24	& 8.4	& 2.27	& 1.84	& 17.1	& 0.108	\\
 A2256	&0.0601	&0.73	& 0.44	& 7.4	& 2.13	& 2.31	& 14.1	& 0.163	\\
 A2319	&0.0564	&0.69	& 0.46	& 9.9	& 2.47	& 4.01	& 21.9	& 0.183	\\
 A3667  &0.0585	&0.51	& 0.25	& 6.5	& 2.00	& 2.40	& 11.6	& 0.206	\\

\hline
\end{tabular}
\end{center}
NOTE.---%
Masses calculated assuming $q_0$=0.0.
\end{table*}

We find that within 1 $h_{50}^{-1}$ Mpc, the gas fraction of the
nearby JF98 subsample is $f_g = (0.186 \pm 0.013) h_{50}^{-3/2}$,
where the standard deviation for a single cluster is 0.048. 
This is consistent with Evrard (1997), who finds
$f_g(r_{500}) = (0.170 \pm 0.008) h_{50}^{-3/2}$ for a sample of
nearby clusters with no luminosity cutoff. 
The observed scatter in $f_g$ is 26\%, in
excellent agreement with simulations (Danos \& Pen 1998). 
The calculated gas fraction of a cluster has the cosmological dependence
$f_g \propto h^{3/2} \Omega _b/\Omega_0$. Since simulations suggest
that $\Omega _b/\Omega_0$ is constant for a given cluster, 
we should be able to
directly measure the redshift evolution of the Hubble constant $h(z)$.
We can derive the cosmological dependence of the cluster gas mass
fraction by 
\begin{equation}
f_g =
M_{gas}/M_{tot} \propto \rho_{gas}r^3/r \propto d_L d_A^{3/2}/d_A =
d_L d_A^{1/2}. 
\end{equation}
Since $d_L d_A^{1/2} =  d_L^{3/2}/(1+z)$, where $d_A$ is the angular
diameter distance, the dependence of gas mass fraction on $q_0$ for any
individual cluster is $f_g(q_0)/f_g(q_0 = 0) = [d_L(q_0)/d_L(q_0 =
0)]^{3/2}$ or 
\begin{eqnarray}
\label{pen}
f_g(q_0) = \Bigl((1-q_0+q_0z+(q_0-1)\sqrt{2q_0z+1})/(q_0^2(z+z^2/2))
\Bigr)^{3/2} \nonumber \\ 
\times f_g(q_0 = 0)
\end{eqnarray}
(Pen 1997, Zombeck 1990) where $f_g(q_0 = 0)$ is 
the gas fraction calculated assuming 
$q_0 = 0$ and $f_g(q_0)$ is the gas fraction calculated assuming a
different value of $q_0$. Since we
assume $q_0 = 0.0$ throughout our analysis, we must derive the correct
value of $q_0$ using the above equation. 

Figure 2 shows the
gas fractions (calculated using $q_0$ = 0.0) of all of the clusters
studied versus redshift along with the predicted apparent evolution in
gas fraction for $q_0$ = 0.5
($\Omega_0 = 1, \Lambda_0 = 0$), $q_0$
= 0.0  ($\Omega_0 = 0, \Lambda_0 = 0$), and $q_0$ = -1.0 ($\Omega_0 =
0, \Lambda_0 = 1$). Errors shown are 90\% confidence limits from the
temperature measurements, which are the dominant source of error. 
The solid squares are clusters noted in JF98 as ``single''
morphological type, whereas the open squares are clusters observed to
contain significant substructure. The scatter in the 
``single'' clusters is significantly smaller than for clusters which are
known to violate the assumptions of hydrostatic equilibrium and
isothermality. 
Because we currently lack adequate spatial 
resolution to make similar analyses of the distant clusters in our
sample, we must 
include the full 25\% scatter in $f_g$ to place meaningful constraints
on $q_0$. Since the nearby sample is not located at $z = 0.0$, we also
apply a correction to the nearby gas fraction assuming $z = 0.06$ (the
mean value) for all of the nearby clusters. 
We define $\chi ^2$ = $\Sigma [f_g^i(q_0, z_i) - f_g(q_0)]^2/
\sigma_i^2$, where $\sigma_i^2$ is the sum in quadrature of the 
90\% confidence limits in temperature (a conservative estimate of the
the 1-$\sigma$ uncertainties in the observations), 
the 7\% uncertainty in the
present gas mass fraction, and the 25\% scatter expected from
simulations. With this definition, 
we find a best fit value of $q_0$ = 0.07 with an acceptable
$\chi ^2$ = 2.94 for 7 degrees of freedom. We calculate 95\%
confidence limits using $\Delta \chi ^2$ = 3.841 and assuming
the errors are Gaussian. This gives a range of $-0.47 < q_0 < 0.67$. 
If we assume the worst-case scenario of an actual evolution of 2\% in
gas fraction due to differences in gas stripping and injection from
galaxies, the limits relax slightly to $-0.51 < q_0 < 0.74$.

\begin{figure} [tb]
\plotone{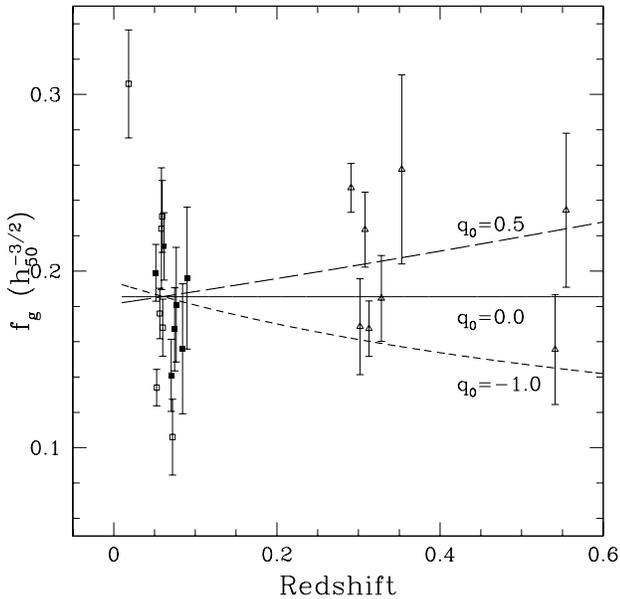}

\figcaption{Observed gas fraction as a function of redshift, calculated
assuming $q_0$ = 0.0. From top to bottom, the lines show the 
apparent evolution of gas fraction predicted for $q_0$ = 0.5
($\Omega_0 = 1, \Lambda_0 = 0$), $q_0$
= 0.0  ($\Omega_0 = 0, \Lambda_0 = 0$), and $q_0$ = -1.0 ($\Omega_0 =
0, \Lambda_0 = 1$). Errors shown are 90\% confidence limits from
temperatures and do not include the 25\% scatter expected from 
simulations.}
\end{figure}

To apply the method of EMN to find the present value of $f_g(r_{500})$, 
we must know the value of $r_{500}$ and the cluster temperatures,
which we take from David et al. (1993). These parameters
and the resulting masses and gas fractions are shown in Table 3. 
We find an average gas fraction of $f_g(r_{500}) = 0.176 \pm
0.011$, where the standard deviation for a single
cluster is 0.043 (the observed scatter is 24\%), which
is again consistent with the gas fraction of Evrard (1997) and the
scatter in the simulations of Danos \& Pen (1998).
Our results support the hypothesis that the inferred gas mass
fractions using the EMN estimator of total mass within $r_{500}$
yields gas mass fractions with
slightly less scatter than those inferred using a
constant radius of 1 $h_{50}^{-1}$ Mpc, 
but the reduction in scatter is not as significant
as in simulations. Since the scatter for nearby clusters was not
significantly reduced, we again assume an intrinsic scatter of 25\% in
the distant cluster gas fractions.
Using the method of EMN, we obtain -0.50 $ < q_0 <$ 0.64 with
95\% confidence (again assuming Gaussian errors). 
The best-fit is at $q_0$ = 0.04, with an acceptable
$\chi ^2 = 3.90$ for 7 degrees of freedom. 
If we assume that the scatter does actually decrease, the limits on
$q_0$ narrow slightly without changing the best fit value. 
Again assuming the worst-case scenario of 2\% actual evolution in gas
fraction due to galaxies, the limits become $-0.53 < q_0 < 0.71$.
If we use Evrard's (1997) value of $f_g(r_{500})=0.170$ for all nearby
clusters (no luminosity selection), we obtain a best fit of $q_0 = 0.16$
with 95\% confidence limits -0.47 $ < q_0 <$ 0.91, so the luminosity
cutoff is not a source of significant systematic effects. 
Figure 4 shows the
variation of $\chi ^2$ with $q_0$ for both methods. The excellent
agreement between the two methods suggests that there are no
significant systematic differences.

\begin{figure} [tb]
\plotone{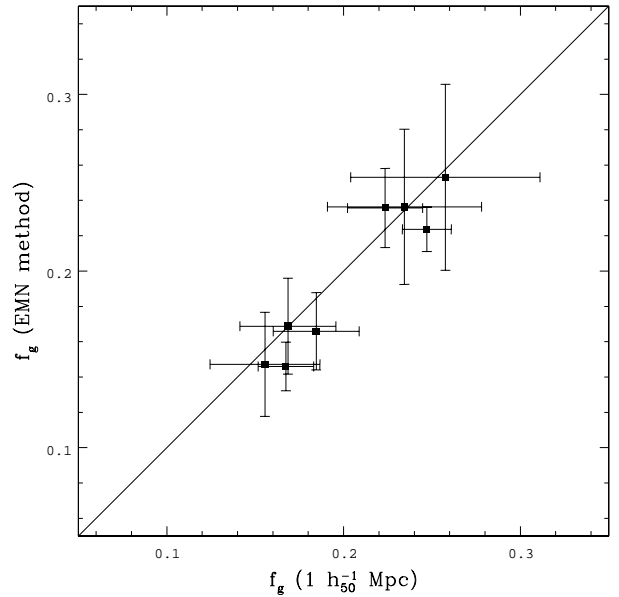}

\figcaption{Gas fractions calculated from usual method ($f_g(1
h_{50}^{-1}$ Mpc)) and using EMN
method to obtain $M_{tot}(r_{500}$) and hence $f_g(r_{500}$). Errors
shown are 90\% confidence limits from temperature measurements.
}

\end{figure}

\begin{figure} [tb]
\plotone{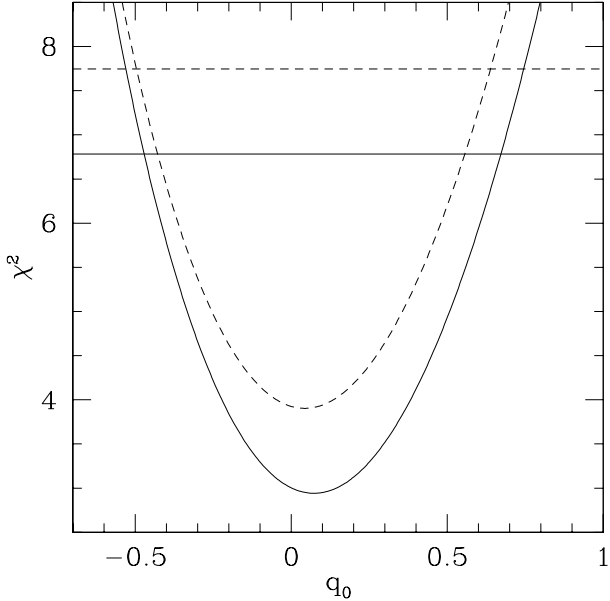}

\figcaption{$\chi ^2$ as a function of $q_0$. The solid line is the
usual method of calculating gas fractions (1 $h_{50}^{-1}$ Mpc) and
the dashed line is using the method of EMN ($r_{500}$). The horizontal
lines show 95\% confidence levels ($\Delta \chi ^2 = 3.841$).
The best fit for the usual method is $q_0$ =
0.07, with 95\% confidence limits of -0.47 $< q_0 <$ 0.67. The EMN method
yields a best fit of 0.04 and limits of -0.50 $< q_0 <$ 0.64.
}
\end{figure}

The low values of $\chi^2$ for the best fits above suggest that our
error estimates are too conservative. A more robust way of
calculating the confidence limits on $q_0$ is to use bootstrap
resampling (Press et al. 1992). We drew $10^5$ artificial data sets
from the actual samples for each of our two methods. For the fixed physical
radius method,  
95\% of the calculated values of $q_0$ lie between -0.18 and 0.66. For
the EMN method, 95\% of the calculated values of $q_0$ lie between
-0.28 and 0.70. These calculations include none of the intrinsic 25\%
scatter from simulations. If we again assume a maximum 2\% actual
evolution in the gas fraction, the limits change to $-0.22 < q_0 <
0.74$ and $-0.32 < q_0 < 0.78$. The agreement with the above results is
encouraging, as are the narrower confidence limits.

\section{CONCLUSIONS}

We have used the apparent change of the cluster gas mass fraction
$f_g  = M_{gas}/M_{tot}$ with redshift together with the assumption of
constant true gas fraction 
to derive constraints on the cosmological deceleration parameter
$q_0$. Our sample is the first ``large'' sample of high redshift
clusters with a common method of analysis, which is important for an
accurate measurement of the apparent change in $f_g$ with redshift. 
We compare
high redshift clusters to a similar sample of nearby clusters,
namely those with $L_X > 10^{45}$ erg s$^{-1}$, though our results are
not strongly affected by this selection. We include the
intrinsic 25\% scatter in $f_g$ predicted by simulations and verified
by observations. We obtain $-0.47 < q_0 < 0.67$
at 95\% confidence using the gas fractions at a constant physical
radius of 1 $h_{50}^{-1}$ Mpc and 
-0.50 $ < q_0 <$ 0.64 with 95\% confidence for gas fractions
evaluated at a constant overdensity radius. The excellent agreement
between these two results suggest that this method to calculate
$q_0$ is quite promising. Bootstrap resampling yields narrower
confidence limits, $-0.18 < q_0 < 0.66$ and $-0.28 < q_0 < 0.70$
respectively for the two methods.
Our results are in good agreement with the upper limit
$q_0<$0.75 from galaxy counts in redshift surveys (Schuecker et
al. 1998) and the lower limit $q_0>-2$ from studies of gravitational
lensing (Park \& Gott 1997). 

We find that the standard closed, flat, and open models for the
universe are all within the formal 95\% confidence limits, as are some
flat $\Omega + \Lambda = 1$ models with non-zero cosmological constant.
We are able to rule out a universe with $\Omega = 0$ and $\Lambda=1$. 
Our results are consistent with observations of high-redshift Type Ia
supernovae (Perlmutter et al. 1997, 
Garnavich et al. 1998, Riess et al. 1998) as well as with gravitational
lensing statistics, which yield $\Lambda < 0.65$ (Kochanek 1996) at
95\% confidence. 

Determining $q_0$ from distant cluster gas fractions has the
significant advantage of being 
independent of $H_0$ and the numerical value of $f_g$. 
Systematic effects could be created by calibration offsets, unknown
selection effects,
non-Gaussian errors, or cooling flows and mergers which are
responsible for much of the intrinsic scatter in the gas fraction.
We have included the expected intrinsic scatter in our calculations,
and the results of Henry (1997) suggest that calibration offsets
between fluxes from {\it Einstein}  and {\it ASCA} are not very
significant. The actual cluster gas mass fractions may increase
slightly with redshift. This would be true if the gas in nearby
clusters has been expelled from the core (e.g., by shock heating)
systematically 
more than in distant clusters. Simulations suggest that this effect is
negligible in both $\Omega = 1$ and $\Omega < 1$ universes.
Alternatively, our results would be biased if the local value
of $H_0$ is significantly different from the global value,
although ``local'' here refers to an average redshift of $z$=0.06.
From Figure 2, we can see that the nearby gas mass fraction has much
less scatter for ``single'' clusters (i.e., ones with no significant
substructure). Thus, constraints on $q_0$ can be greatly
improved not just by increasing the sample size but by improving the
analysis of individual distant clusters so we can differentiate
between clusters with and without substructure and properly measure
temperature distributions. In conclusion, although our current results
do not improve significantly the determination of $q_0$, we have shown
that with the assumption of constant gas mass fractions, future
detailed X-ray observations with {\mbox AXAF} and {\mbox XMM} can
provide useful constraints on $q_0$.

\acknowledgments
This research has made use of data obtained through the
HEASARC Online Service, provided by the NASA/Goddard
Space Flight Center. K. Rines would like to thank the
National Science Foundation for creating the REU program at the
Smithsonian Astrophysical Observatory where this work began. 
K. Rines thanks L. David and A. Vikhlinin for their
assistance. Several of the clusters in this study were observed as
part of the "Northern ROSAT All-Sky Atlas Cluster Survey" by
John Huchra, Rich Burg, Brian McLean and Hans B\"{o}hringer. John
Huchra and Leon VanSpeybroeck provided useful comments prior to
publication. The presentation of this paper was significantly improved
by the comments of an anonymous referee. W. Forman and C. Jones
acknowledge support from the 
Smithsonian Institution and NASA contract NAS8-39073.


\end{document}